\documentclass[%
    reprint,
    superscriptaddress,
     amsmath,amssymb,
      aps,
  ]{revtex4-2}
  \usepackage{xcolor}
  \usepackage{graphicx}
  \usepackage{dcolumn}
  \usepackage{physics}
  \usepackage{bm}
  \usepackage{tikz}
  \usepackage[normalem]{ulem} 

\begin{document}
\title[Charge Qubit in 2D Bilayer Materials]{Dynamics of van der Waals Charge Qubit in 2D Bilayers: \textit{Ab initio} Quantum Transport and Qubit Measurement}

\author{Jiang Cao}
\affiliation{Integrated Systems Laboratory, ETH Z\"urich, 8092 Z\"urich, Switzerland}

\author{Guido Gandus}
\affiliation{Integrated Systems Laboratory, ETH Z\"urich, 8092 Z\"urich, Switzerland}
\affiliation{Empa, Swiss Federal Laboratories for Materials Science and Technology, \"Uberlandstrasse 129, 8600 D\"ubendorf, Switzerland}

\author{Tarun Agarwal}
\affiliation{Department of Electrical Engineering, IIT Gandhinagar, Gandhinagar, India}

\author{Mathieu Luisier}
\affiliation{Integrated Systems Laboratory, ETH Z\"urich, 8092 Z\"urich, Switzerland}

\author{Youseung Lee}
\email{Corresponding author: youseung.lee@iis.ee.ethz.ch}
\affiliation{Integrated Systems Laboratory, ETH Z\"urich, 8092 Z\"urich, Switzerland}

\begin{abstract}
A van der Waals (vdW) charge qubit, electrostatically confined within two-dimensional (2D) vdW materials, is proposed as building block of future quantum computers. Its characteristics are systematically evaluated with respect to its two-level anti-crossing energy difference ($\Delta$). Bilayer graphene ($\Delta$ $\approx$ 0) and a vdW heterostructure ($\Delta$ $\gg$ 0) are used as representative examples. Their tunable electronic properties with
an external electric field define the state of the charge qubit. By combining density functional theory and quantum transport
calculations, we highlight the optimal qubit operation conditions based on charge stability and energy-level diagrams.
Moreover, a single-electron transistor (SET) design based on trilayer vdW heterostructures capacitively coupled to the charge qubit is introduced as measurement setup with low decoherence and improved measurement properties.
It is found that a $\Delta$ greater than 20~meV results in a rapid mixing of the qubit states, which leads to a lower measurement quantity, i.e. contrast and conductance.
With properly optimized designs, qubit architectures relying on 2D vdW structures could be integrated into an all-electronic quantum computing platform.
\end{abstract}

\keywords{quantum computing, charge qubit, two-dimensional materials, bilayer graphene, quantum transport}

\maketitle

\section{Introduction}
\label{sec:introduction}
\par Emerging nanoelectronic devices are widely seen as key enablers of future quantum computers that will be able to solve intractable problems for classical machines. A wide range of qubit platforms have already been experimentally demonstrated, e.g. charge, spin, superconducting, and trapped-ion qubits \cite{DQD,SpinQubit,SuperconductingQubit,TrapIonQubit}. In all these approaches, a large number of physical qubits are required to realize one single logical qubit \cite{blueprintQC}. The ideal qubit structure should therefore be scalable and integrable with the existing CMOS technology to build a large-scale quantum computer. 

\par Quantum dots (QD) offer a great platform to generate CMOS-compatible qubits based on the spin and/or electronic orbital degrees of freedom of electrons \cite{Exp1,Exp2,QDNanoLett}. For example, silicon spin qubits show very long coherence times up to milliseconds at low temperature~\cite{SpinQubit}. However, they are rather challenging to handle due to the rapid flip-flop of single electron spins. Charge qubits, on the other hand, can simplify the read-out of qubits thanks to well-developed technologies such as single-electron transistors (SET).
Moreover, these qubits can be directly operated by electrodes up to gigahertz rates. 

\begin{figure*}[!ht]
\centerline{\includegraphics[width=0.9\linewidth]{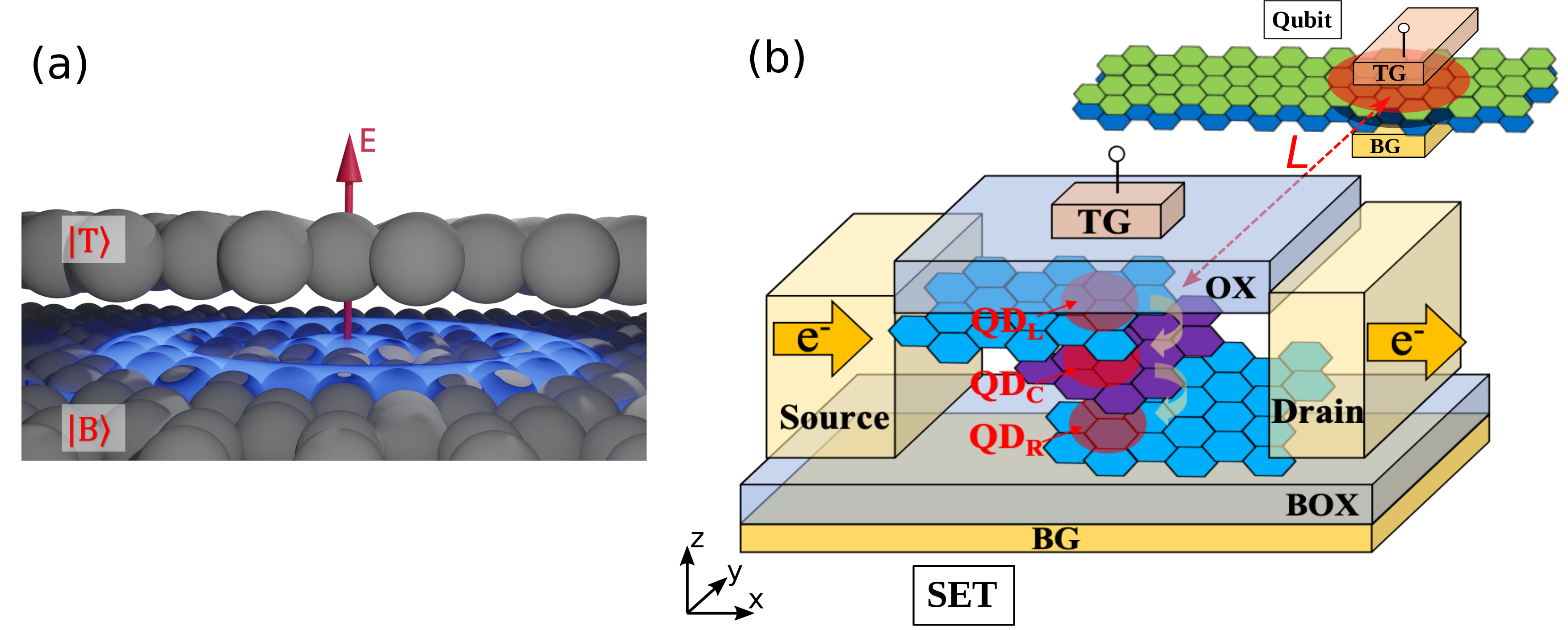}}
\caption{
(a) Sketch of a charge qubit system implemented on a 2D bilayer material to which a vertical electric field $\vec{E}$ is applied. The wave functions localized in the bottom and top layers are labeled $\ket{B}$ and $\ket{T}$, respectively.  
(b) Schematic of the proposed qubit measurement setup where three QDs labeled QD$_{L,C,R}$ are confined inside a vdW trilayer MoS$_2$-WSe$_2$-MoS$_2$ heterostructure controlled by a top (TG) and bottom gate (BG) and connected to a source and drain contact forming altogether a single-electron transistor (SET). The trilayer MoS$_2$-WSe$_2$-MoS$_2$ SET is electrostaticlly coupled to the charge qubit located on the BLG. The distance between the SET and qubit is $L$.
}
\label{fig:1}
\end{figure*} 

\par Two-dimensional (2D) materials represent a promising alternative to conventional semiconductors to host qubits owing to the wide tunability of their electronic properties and the manyfold quantum phenomena that emerge in the atomically thin limit such as massless Dirac fermions~\cite{Novoselov2005}, anyons~\cite{Nayak2008}, or the quantum spin Hall effect~\cite{Xiaofeng2014}. 
In addition, the van der Waals (vdW) stacking of 2D materials provides versatility in growth, functionalization, and heterostructure integration, which facilitates the processing of complex systems. AB-stacked bilayer graphene (BLG), a vdW homojunction, is an excellent example: its band gap can be gradually modified by applying an external electric field~\cite{BLG_PRL_Falko,MCCANN2007110,AOKI2007123,BLG_PRL_Novoselov,BLG_Nature,BLG_PRL_Mak,blg}. 
Moreover, the electric field can modify the quantum superposition of wave functions localized in different layers, affecting both the conduction and valence bands of BLG. 
The same effect can be obtained in other vdW structures as well if the band edges of the individual 2D materials that constitute the stack get nearly aligned on an absolute energy scale~\cite{Koda2018}.
Based on this observation, vdW structures are attractive to implement charge qubits capable of switching between two states with the help of a vertical electric field~\cite{ChargeQubit}. However, for vdW charge qubits to become a viable option, an all-electronic platform that can host multiple QDs should be available. The coupling between neighboring QDs should be tunable, while it should be possible to measure the qubit state with low decoherence.

\par In this work, we propose a platform for vdW charge qubits that takes advantage of quantum superposition of the top and bottom charge states formed by vdW bilayers, as illustrated in Fig.~\ref{fig:1}(a), and we analyze its potential through \textit{ab initio} device simulations. Differently from the conventional charge qubits that require two QDs, e.g. the left and right QDs, a single QD of bilayer acts as an individual qubit. As a promising candidate, we showcase the functionality of a BLG vdW charge qubit system. The role of two-level anti-crossing energies $\Delta$ is analyzed for two exemplary vdW materials that illustrate the extreme cases of $\Delta$: BLG ($\Delta \approx 0$) and ZrS$_2$-HfS$_2$ ($\Delta = 0.174$~eV). As a first step, density functional theory (DFT) and quantum transport calculations are combined to study the charge stability diagram and energy-level spectrum of multiple vdW charge qubits with tunable coupling, formed on an electrostatically confined BLG device. 
It should be pointed out that directly coupled read-out leads, as originally proposed in the paper of Lucatto \textit{et al.}~\cite{ChargeQubit}, can be prominent sources of decoherence due to the finite tunnel coupling of the reservoirs to the charge qubit. As a consequence, it might be experimentally difficult to measure the qubit state. To overcome this issue, we designed a novel measurement setup using triple-QD SETs based on a trilayer vdW heterostructure electrostatically coupled to our vdW charge qubits, as depicted in Fig.~\ref{fig:1}(b). Here, its influence will be simulated through the Lindblad master equation with the QuTiP package \cite{Qutip} using a phenomenological model including the important physics of decoherence.
Our results highlight that 2D vdW structures can provide highly integrable all-electronic universal qubits. Moreover, our findings indicate that devices with $\Delta \gg$ 20 meV pose specific design challenges related to the material combination and geometrical parameters. Hence, a careful selection of the $\Delta$ value is necessary to design properly working vdW charge qubits.

\section{vdW charge qubit manipulation}
\label{vdWQubit}
\begin{figure*}[!htp]
	\centering
        \includegraphics[width=0.9\linewidth]{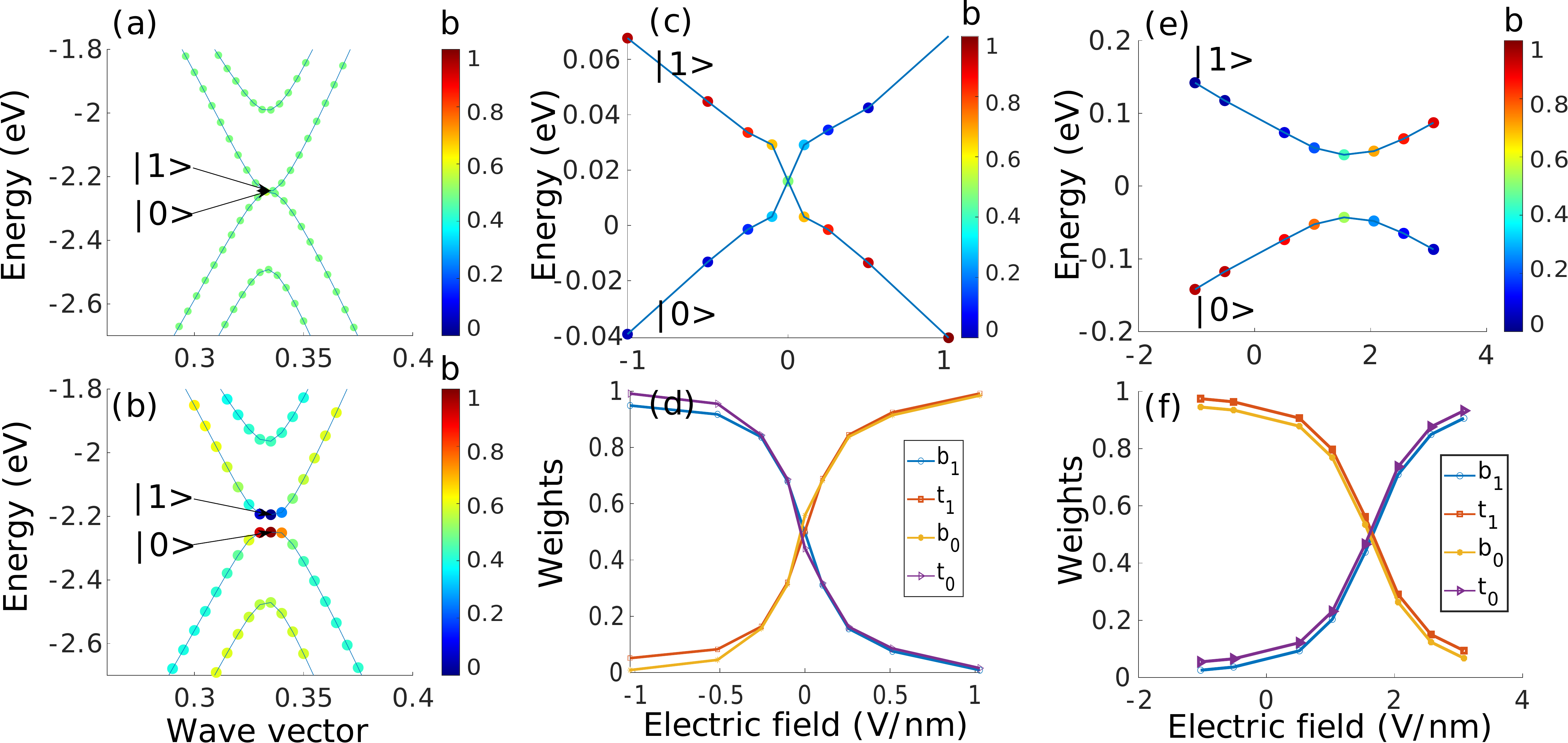}
		\caption[]{Electronic bandstructure of BLG (a) without and (b) with an electric field of 0.5~V/nm. The color of the different states indicates the $\abs{b_{0,1}}^2$ weights, which result from the projection of the atomic wave functions onto the bottom graphene layer. The valence band maximum and conduction band minimum define the $\ket{0}$ and $\ket{1}$ states. (c) Energy-level vs. electric field for the $\ket{0}$ and $\ket{1}$ states in BLG. (d) Corresponding $\abs{b_{0,1}}^2$ and $\abs{t_{0,1}}^2$ weights vs. electric field. (e) Same as (c) but for ZrS$_2$-HfS$_2$. (f) Same as (d) but for ZrS$_2$-HfS$_2$.}
	\label{fig:2}
\end{figure*}

We first examine the effect of a vertical electric field on the selected 2D vdW bilayer materials, BLG and ZrS$_2$-HfS$_2$, and show that the electric-field externally applied to the bilayers can be used to control the wave function localized in the top or bottom layer to construct the qubit states.
The BLG  band structure without and with electric field is shown in Figs.~\ref{fig:2}(a) and \ref{fig:2}(b), respectively. The detailed DFT simulation procedure that was used to compute those quantities can be found in Appendix~\ref{Methodology1}. The  conduction band minimum (CBM) and valence band maximum (VBM) are chosen as the two-level qubit states ($\ket{0}$ and $\ket{1}$) in the case of BLG, while the first and second lowest conduction bands of the ZrS$_2$-HfS$_2$ heterostructure play the same role. 
One importance for vdW bilayers to properly act as qubits is the existence of two bands lying quite close to each other and that can be easily inverted by applying a practical electric field.

\par In the BLG system without electric field $\vec{E}$, the $\ket{0}$ and $\ket{1}$ states are made of exactly equal contributions from each graphene layer due to the inversion symmetry of this material. A bandgap can be opened by applying a symmetry-breaking vertical $\vec{E}$. 
The vdW charge qubit states, $\ket{0}$ and $\ket{1}$, are constructed by combining the wave functions localized in the top and bottom layers, $\ket{T}$ and $\ket{B}$, respectively.

\par By applying a positive $\vec{E}$ along the $+z$ direction, as in Fig.~\ref{fig:1}(a), the bands of the bottom graphene layer are lowered in energy, 
which increases the $\ket{B}$ character of $\ket{0}$, while $\ket{1}$ becomes more
localized in the top graphene layer. The opposite behavior happens with $\vec{E}$ pointing along the $-z$ direction. 
The contribution of each layer to the Bloch wave function of $\ket{0}$ and $\ket{1}$ strongly depends on the vertical electric field. An electron wave function $\ket{\psi(\vec{E})}$ can be written as the superposition of the wave functions $\ket{T}$ and $\ket{B}$ for a given $\vec{E}$
\begin{equation} \label{eq:state}
    \ket{\psi(\vec{E})} = t_\psi(\vec{E}) \ket{T} + b_\psi(\vec{E}) \ket{B},
\end{equation}
where $t_\psi(\vec{E})$ and $b_\psi(\vec{E})$ are complex coefficients depending on $\vec{E}$. Due to the vdW gap between the two layers, the $\ket{T}$  and $\ket{B}$ states have a relatively small overlap and form a quasi-orthogonal basis. The $t_\psi(\vec{E})$ and $b_\psi(\vec{E})$ coefficients satisfy the normalization condition $\abs{t_\psi}^2+\abs{b_\psi}^2=1$.

\par Figure~\ref{fig:2}(c) shows the energy level of the $\ket{0}$ and $\ket{1}$ states and the $\ket{B}$ contribution to them as a function of $\vec{E}$. In Fig.~\ref{fig:2}(d) the $ \abs{t_{0,1}}^2$ and $\abs{b_{0,1}}^2$ weights of the $\ket{0}$ and $\ket{1}$ states are plotted, highlighting a complementary relation between them. Here, we note that by tuning $\vec{E}$ we can readily address any coherent superposition of the basic quantum states $\ket{B}$ and $\ket{T}$. The state $\ket{\psi}$ can be equivalently described by a Bloch vector in the standard Bloch sphere representation, where the coefficients are given by a polar $\theta$ and azimuthal angle $\phi$ such that $t_\psi=\cos(\theta/2)$ and $b_\psi=\sin(\theta/2) e^{i\phi}$.
Moreover, under a reasonably strong positive/negative electric field, the $\ket{0}$ and $\ket{1}$ states can be localized in either the top or bottom layer, thus enabling the ``must-have'' qubit initialization process.

\begin{figure*}[htp]
	\centering
        \includegraphics[width=0.9\linewidth]{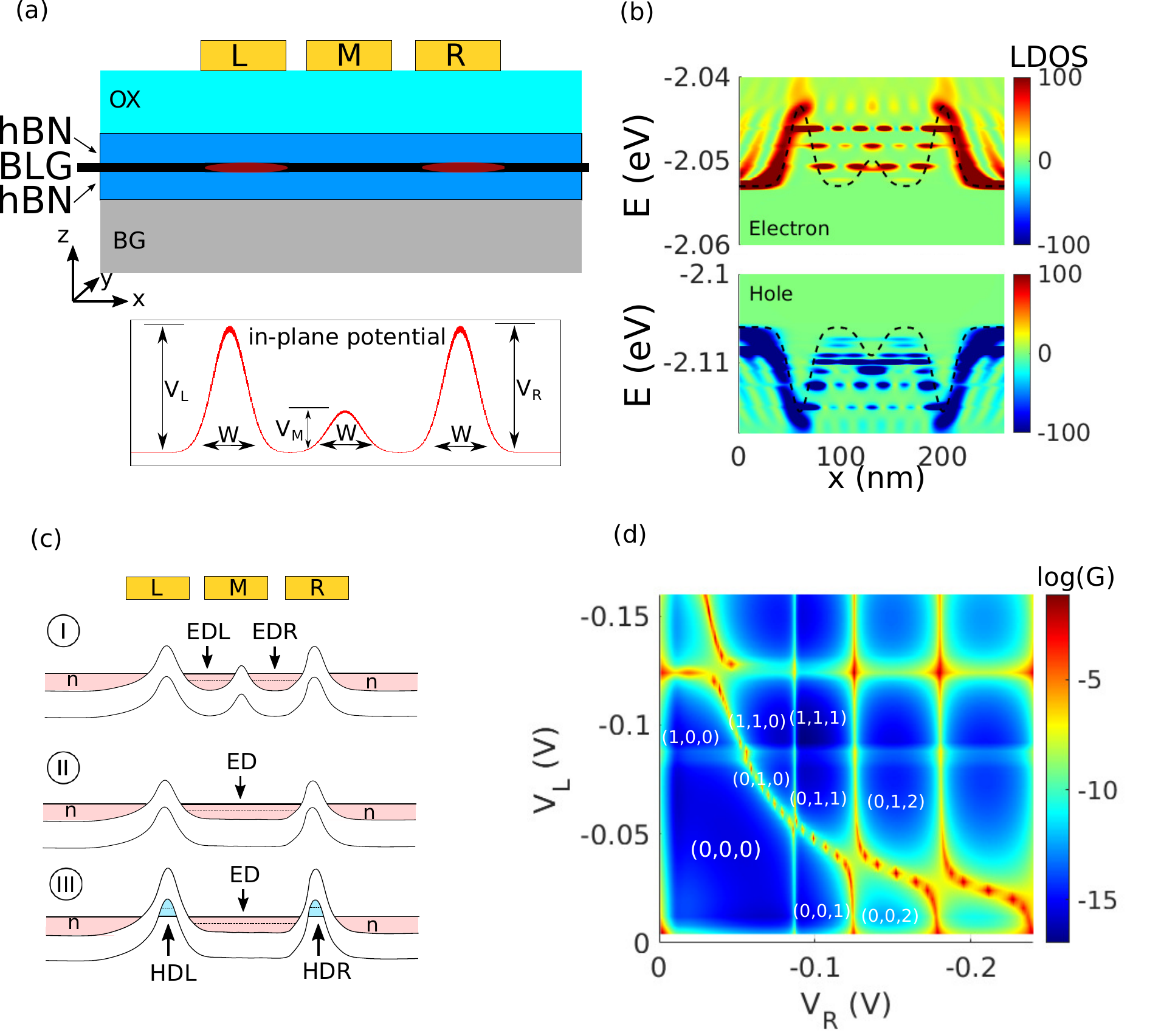}
		\caption[]{
		(a) Simulation domain corresponding to a bilayer graphene (BLG) with a total length of 260~nm along the transport direction $x$ and a width of 2.1~nm along $y$ for a total of 42400 atoms, with the periodic boundary condition along $y$. A vertical electric field $\Vec{E}$ acts on the BLG and is applied through a back gate (BG) contact. The BLG is embedded between two $h$BN layers. The red regions under L and R indicate the position of the electrostatically confined quantum dots. The electrostatic potential is further controlled by three top gates labeled L, M, and R. They are separated from the BLG by a SiO$_2$ oxide and a $h$BN layer. The bottom inset shows a typical potential profile obtained by superposing Gaussian functions induced by each top electrode.
		(b) Local density-of-states of the electron and hole populations. Positive (negative) values refer to the top (bottom) layer, respectively. $\vec{E}=0.5$~V/nm, $V_{\rm M}=\pm4$~mV, and $V_{\rm L}=V_{\rm R}=\pm0.01$~V where + stands for hole while - for electron. The dashed lines represent the conduction and valence band edges. 
		(c) Schematics of the band profile along the BLG sheet illustrating the formation of dots: (I) using the M gate, two n-doped islands form the desired electron dots (ED) with a tunneling barrier between them; (II) at $V_{\rm M}=0$, one large ED is formed; (III) at larger $V_{\rm L}$ and $V_{\rm R}$, two additional hole dots (HD) are created under these gates. 
		(d) Charge stability diagram: logarithmic conductance at a drain-to-source voltage $V_{\rm DS}$=0.1~mV, Fermi level $E_F$=-2.0525~eV, $\vec{E}=0.5$~V/nm, $V_{\rm M}$=0 V and $k_B T$=1~$\mu$eV as a function of the L and R gate voltages. The equilibrium charge number in HDL, ED and HDR are depicted as (N$_0$,N$_1$,N$_2$).}
	\label{fig:3}
\end{figure*}

\par A similar behavior is observed for the first and second lowest conduction band states of ZrS$_2$-HfS$_2$, as depicted in Fig.~\ref{fig:2}(e) and (f). Due to the energy offset in the natural band alignment of the vdW heterostructure, a non-zero $\vec{E}_{ac}$ is required to reach the crossing point where the layer contributions are equal. At this point, a so-called anti-crossing energy $\Delta$ can be introduced. The latter is then a material parameter and determined as the difference between the energies of the $\ket{1}$ and $\ket{0}$ eigenstates~\cite{DQD}. Table I summaries the values of $\Delta$ for the considered vdW structures. For BLG, we note that  $\Delta \gtrsim 20$~$\mu$eV was observed in previous experiments~\cite{AOKI2007123,Konschuh2012,Stampfer2021} due to the presence of symmetry-breaking factors even in the absence of an electric field, for example spin-orbit coupling, surrounding substrate atoms, or local strain.

\par By comparing our results to the ZrSe$_2$-SnSe$_2$ vdW heterostructure investigated in the original paper of Lucatto \textit{et al.}~\cite{ChargeQubit}, we notice that the Bloch vector of the BLG and ZrS$_2$-HfS$_2$ systems covers a larger $\theta$ interval for electric fields within $\pm$3~V/nm. For BLG, the full range of $\theta$ can be achieved with a small electric field of less than 1~V/nm, thus making BLG the most technologically appealing candidate.

A quantum control technique over the angle $\phi$ (\textit{i.e.} relative phase of the wave functions) applicable to the vdW charge qubits has been proposed in Ref.~\onlinecite{LZS}. It relies on the Landau-Zener-St\"uckelberg (LZS) interference when the system non-adiabatically sweeps through the intersection point. Based on this technique, general rotations in the Bloch sphere and universal single qubit operations have already been experimentally demonstrated for Si and GaAs DQD charge qubits~\cite{DQD,LZS}.

\section{Quantum transport simulations}
\label{QT}
\par To create a clear two-level system with a $\ket{0}$ and $\ket{1}$ state, the QDs in the vdW bilayers must be formed electrically through gate electrodes that induce a local confinement, as illustrated in Fig.~\ref{fig:3}(a). In the proposed device setup, a bottom gate uniformly acts on the BLG encapsulated within two hBN dielectric layers to electrostatically control the Fermi level. A transport channel can be defined through the BLG. Separated by an oxide, three top gates labeled L, M, and R are placed. By applying different potentials to the bottom and top gates, the vertical electric field acting on the BLG can be precisely controlled. In addition, the potential landscape can be modified by tuning the top L, M, and R gates. The simulated device structure measures 260~nm along the transport direction, possesses zigzag edges, and is made of 42400 atoms in total. The three top gates are 10~nm long and separated by 50~nm. Charge transport is solved with a quantum transport simulator~\cite{QT} based on the \textit{ab initio} Non-equilibrium Green's function (NEGF) formalism, as described in Appendix~\ref{Methodology1}. In this study, the potential profiles are approximated as Gaussian functions that are plotted in the bottom inset of Fig.~\ref{fig:3}(a), Poisson's equation being not self-consistently solved due to the large computational burden associated with the consideration of thousands of gate voltage configurations.

\par Figure~\ref{fig:3}(b) shows that by applying $\vec{E}=0.5$~V/nm and by properly adjusting the applied top gate voltages both electron and hole QDs can be generated in BLG, without spurious bandgap states and with a strong localization of the $\ket{0}$ and $\ket{1}$ states in the top and bottom graphene layer. The projected local density-of-states (LDOS) confirm that the qubit state defined in Fig.~\ref{fig:2} is not altered by the confinement potential. By applying an appropriate bias to the top M  gate (or plunger gate), a gradual transition from two QDs hosting electrons or holes to a single larger QD can be achieved, thus providing a way to tune the interdot tunnel coupling. This transition occurs between regime I and II in Fig.~\ref{fig:3}(c).
The charge stability is demonstrated by the distinct resonances in the conductance map as a function of the gate voltages applied to the L and R electrodes, $V_{\rm L,R}$, setting $V_{\rm M}=0$. $V_{\rm L,R}$ creates two potential barriers that confine an electron dot (ED) in the middle and that tune its energy levels. Each time an energy level crosses the energy window defined by the Fermi energies of the source and drain contacts, a resonance peak can be seen in the conductance map (see Fig.~\ref{fig:3}(d)). In addition to creating n and p-type QDs in the central region, the electrostatic potential can also be shaped to form ambipolar QDs through the application of large enough L and R gate voltages, as highlighted by regime III in Fig.~\ref{fig:3}(c). The L/R gates closely control the energy levels in the hole dot (HDL/HDR) underneath it. Whenever the energy levels of the ED and HDL/HDR cross the energy window of interest, a crossing of two lines of maximum conductance can be observed in Fig.~\ref{fig:3}(d). We note that the observed high LDOS on the QD boundaries, which originated from the coupling to the continuum of states in the left and right contacts due to the NEGF open boundary condition, is expected to be reduced in a fully self-consistent Schr\"odinger-Poisson solution. However, we do not expect this to alter qualitatively our result because the conductance is limited by the electron states inside the quantum dot rather than in the boundaries.

\begin{figure}[htp]
	\centering
        \includegraphics[width=8cm]{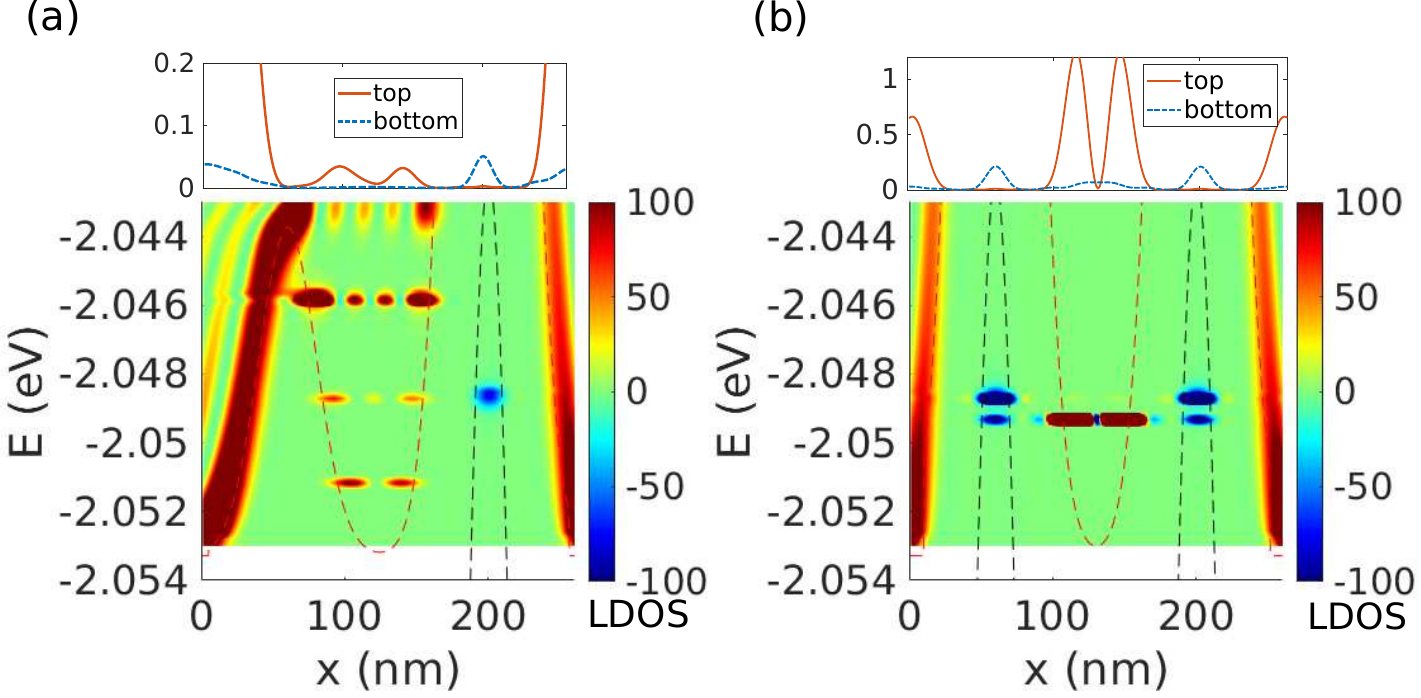}
		\caption[]{Local density-of-states of the electron and hole population of BLG at a positive electric field of 0.5~V/nm.  Positive (negative) values refer to the  top (bottom) layer. (a) At gate voltages $V_M$=0~V, $V_R$=-0.06~V, and  $V_L$=-0.01~V. (b) Same as (a), but at $V_L$=-0.06 V. The dashed lines represent the conduction and valence band edges. The corresponding charge density distributions in the top and bottom layers with $E_F$=-2.048~eV are plotted in the upper panels.}
	\label{fig:4}
\end{figure}

\par In order to investigate the system scalability towards a highly integrable qubit platform, we also show the posibility of creating multiple qubits within one BLG device in Fig.~\ref{fig:4}. Since the electron and hole dots possess well-separated LDOS and charge densities localized in the top and bottom layers, they can be exploited for hosting $\ket{1}$ and $\ket{0}$ qubit states. It is worth noting that between the QDs, the Fermi energy lies inside the band gap, which
leads to a finite region of zero charge density and a tunneling barrier. Figure~\ref{fig:4}(a) shows an n-p double-QD, while Fig.~\ref{fig:4}(b) presents a p-n-p triple-QD formed in the channel. 
A chain of QDs of $\ket{1}$ and $\ket{0}$ can be constructed via $N$ top gates, thus promising easily scalable multi-qubit systems \cite{QDNanoLett}. 
Contrary to the traditional charge qubits where two QDs are required to form a single charge qubit~\cite{DQD,Kim2015}, 
a single vdW-bilayer QD is sufficient to form one qubit because the wave function in each dot is a quantum superposition of $\ket{T}$  and $\ket{B}$. 
Moreover, a single global bottom gate electrode reduces the number of gate electrodes, which greatly simplifies the geometrical complexity and increases the integration density of qubits.
Coupling remote charge qubits has been experimentally demonstrated for graphene~\cite{Deng2015,Deng2015a} using microwave resonators, offering the possibility to entangle more-than-nearest-neighbor QDs. While the previous experiments show a viable path to implement a multi-qubit coupling scheme for the vdW-bilayer QDs, this remains a challenging task that will need detailed investigations in the future. They go far beyond the scope of our present work.

\section{Time-dependent qubit measurement}
\label{QubitReadout}
\par After demonstrating the qubit operation conditions and obtaining  charge stability for multiple QDs, the procedure to read out the qubit state is now discussed. For time-dependent simulations, the
qubit is modeled by a two-level Hamiltonian constructed from the DFT bandstructure of 2D bilayers
under different vertical electric fields
\begin{equation}
\begin{split}
    \hat{H}_{\rm Q}(\vec{E}) = & \epsilon_T(\vec{E}) \ket{T}\bra{T} + \epsilon_B(\vec{E}) \ket{B}\bra{B}\\  
         & - \frac{\Delta}{2} (\ket{T}\bra{B} + \ket{B}\bra{T}) ,
\end{split}
\end{equation}
where $\epsilon_{T}(\vec{E})$ and $\epsilon_{B}(\vec{E})$ are the on-site energies of the top and bottom layer
that depend linearly on $\vec{E}$, and $\Delta$ is the anti-crossing energy.
These values are obtained by fitting the VBM and CBM energy levels for varying $\vec{E}$ (see Fig.~\ref{fig:2}) to the eigenvalues of the quadratic Hamiltonian.

\begin{figure}[!htp]
	\centering
        \includegraphics[width=8cm]{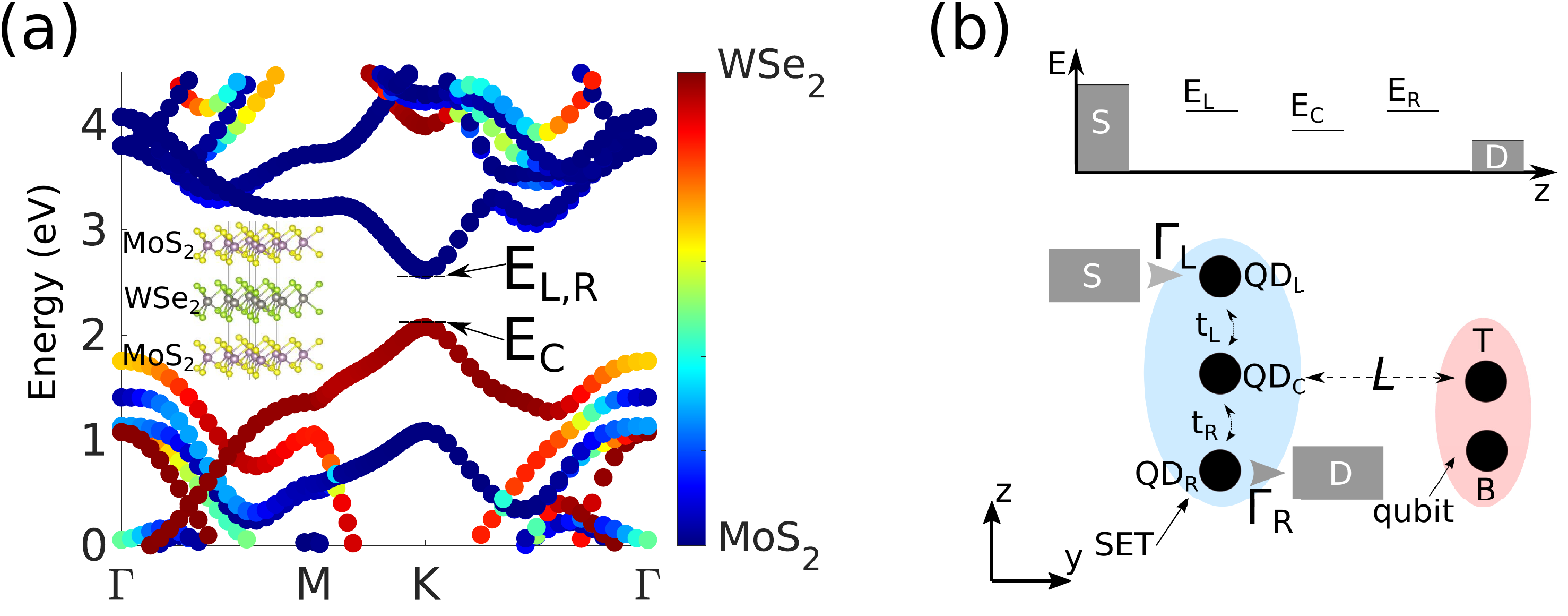}
		\caption[]{(a) Electronic bandstructure of the trilayer vdW heterostructure MoS$_2$-WSe$_2$-MoS$_2$ for the SET measuring setup. The color of the states indicates the contribution from each individual MoS$_2$ and WSe$_2$ monolayer to the overall bandstructure. The inset represents the simulated vdW structure. (b) Conceptual representation of the whole system consisting of one vdW qubit (red circled) and a SET (blue circled) viewed from the y-z cross-section in Fig.~\ref{fig:1}(b). The relevant energy levels of QD$_{L,C,R}$ are $E_{L,C,R}$, respectively. They are shown in the top sub-plot. The tunneling probabilities between the source, QD$_{L,C,R}$ and the drain are $\Gamma_L$, $t_{L,R}$, and $\Gamma_R$, respectively. The SET is coupled through Coulomb interaction to the charge qubit located within a bilayer system at a distance $L$.}
	\label{fig:5}
\end{figure}

A SET formed by a linear array of three QDs capacitively coupled to the qubit is chosen for the
measurement, as suggested in Ref.~\cite{SET} and shown in Fig.~\ref{fig:5}(b). Such a configuration
has an enhanced readout sensitivity as compared to the more common single QD SET. The designed SET is described by the following Hamiltonian operator,
\begin{equation}
\begin{split}
\hat{H}_{\rm SET} =& E_L \ket{L}\bra{L} + E_R \ket{R}\bra{R} + E_C \ket{C}\bra{C}\\
    & - t_L (\ket{C}\bra{L} + \ket{L}\bra{C})\\
    & - t_R (\ket{C}\bra{R} + \ket{R}\bra{C})\\
    & + \sum_{i,j=L,C,R} W_{i,j} \ket{i,j} \bra{i,j},
\end{split}
\end{equation}

where $\ket{L}$, $\ket{R}$, and $\ket{C}$ denote the single-electron states  when the electron is located in QD$_{\rm L}$, QD$_{\rm R}$, and QD$_{\rm C}$, respectively, and $E_{\rm L}$, $E_{\rm R}$, and $E_{\rm C}$ for the on-site energies; $t_L$ and $t_R$ indicate the tunneling probabilities between the QDs; $\ket{L,R}$, $\ket{L,C}$ and $\ket{C,R}$ are the two-electron states, while $W_{i,j}$ represents the Coulomb interaction between dots $i$ and $j$. 
The electron density in the QD is assumed as a point charge. In this case the energies of the electrostatic interaction have the following values  $W_{i,j} = 2 {\rm Ry} \times a_B / d_{i,j} $, where $d_{i,j}$ is the distance between the centers of the QDs $i$ and $j$, $a_B=0.52\times 10^{-10}$~m  is the Bohr radius, Ry = 13.6~eV is the Rydberg energy. This approximation works well for small-sized charge dots as the ones considered in this work~\cite{SET}.
The SET is only coupled to the qubit $via$ the Coulomb interactions without any direct tunneling,
\begin{equation}
\hat{H}_{\rm Q-SET} = \sum_{i=T,B; j=L,C,R} W_{i,j} \ket{i,j} \bra{i,j}.
\end{equation}
The time evolution of the density matrix operator $\hat{\rho}$ corresponding to the qubit-SET system was 
simulated by solving the Lindblad equation \cite{SET,Linblad1}. More details are given in Appendix~\ref{Methodology2}.

The optimal condition for the aforementioned triple-QD SET requires first that the
energetic detunings between the external (QD$_{\rm L}$, QD$_{\rm R}$) and the central (QD$_{\rm C}$) dots, $\delta_{\rm L,R}=E_{\rm L,R} - E_{\rm C}$, are equivalent. Second, the electrostatic couplings $W$ between the SET QDs and the qubit should be symmetric (asymmetric) for the electron in the top (bottom) layer, respectively. More explicitly, $W_{T,L}=W_{T,R}$ and $W_{B,L} \neq W_{B,R}$.
We therefore propose a vdW heterostructure made of three 2D materials (MoS$_2$-WSe$_2$-MoS$_2$) as SET design, as illustrated in Fig.~\ref{fig:1}(b).

\par The projected bandstructure in Fig.~\ref{fig:5}(a) reveals  that  the  CBM  and  VBM  are located within one of the 2D layers. The degenerate CBM of the top and  bottom  MoS$_2$ layers  define  the  energies  of  the  external QDs  L  and  R,  and  the  VBM  of  WSe$_2$ realizes  the  central QD  C.  The  direct  bandgap  at  K in the Brillouin Zone  corresponds to the  symmetric  detuning $\delta_{\rm L,R}$. The vertical alignment of the top qubit layer with WSe$_2$ guarantees that the presence of an electron in this layer causes a symmetric renormalization of the energies in the SET. In contrast, an electron in the qubit bottom layer creates an imbalance of detunings and causes the current to cease.
The triple-QD in the vdW heterostructure is formed electrically by a top gate electrode that induces a local confinement. 
To asses the impact of electron-phonon and electron-electron interactions on the measurement sensitivity of the coupled qubit and SET systems, we included the relaxation and dephasing rates $\gamma_{rel}$ and $\gamma_{dep}$, accounting for those scattering processes, respectively. The corresponding collapse operator in the Lindblad equation is constructed as
\begin{equation} \label{eq:c_ops}
\begin{split}
\mathcal{C} =& \gamma_{rel} [  \mathcal{D} (\ket{C} \bra{L}) + \mathcal{D} (\ket{C} \bra{R}) ], \\
+& \gamma_{dep} [  \mathcal{D} (\ket{L} \bra{L} - \ket{C} \bra{C})  
            +  \mathcal{D} (\ket{R} \bra{R} - \ket{C} \bra{C}) \\
            +&  \mathcal{D} (\ket{T} \bra{T} - \ket{B} \bra{B}) ],
\end{split}
\end{equation}
where $\mathcal{D}(\hat{O}) \equiv \hat{O} \hat{\rho} \hat{O}^\dagger- \frac{1}{2}[\hat{O}^\dagger \hat{O},\hat{\rho}]$.

\begin{figure}[htp]
	\centering
        \includegraphics[width=\columnwidth]{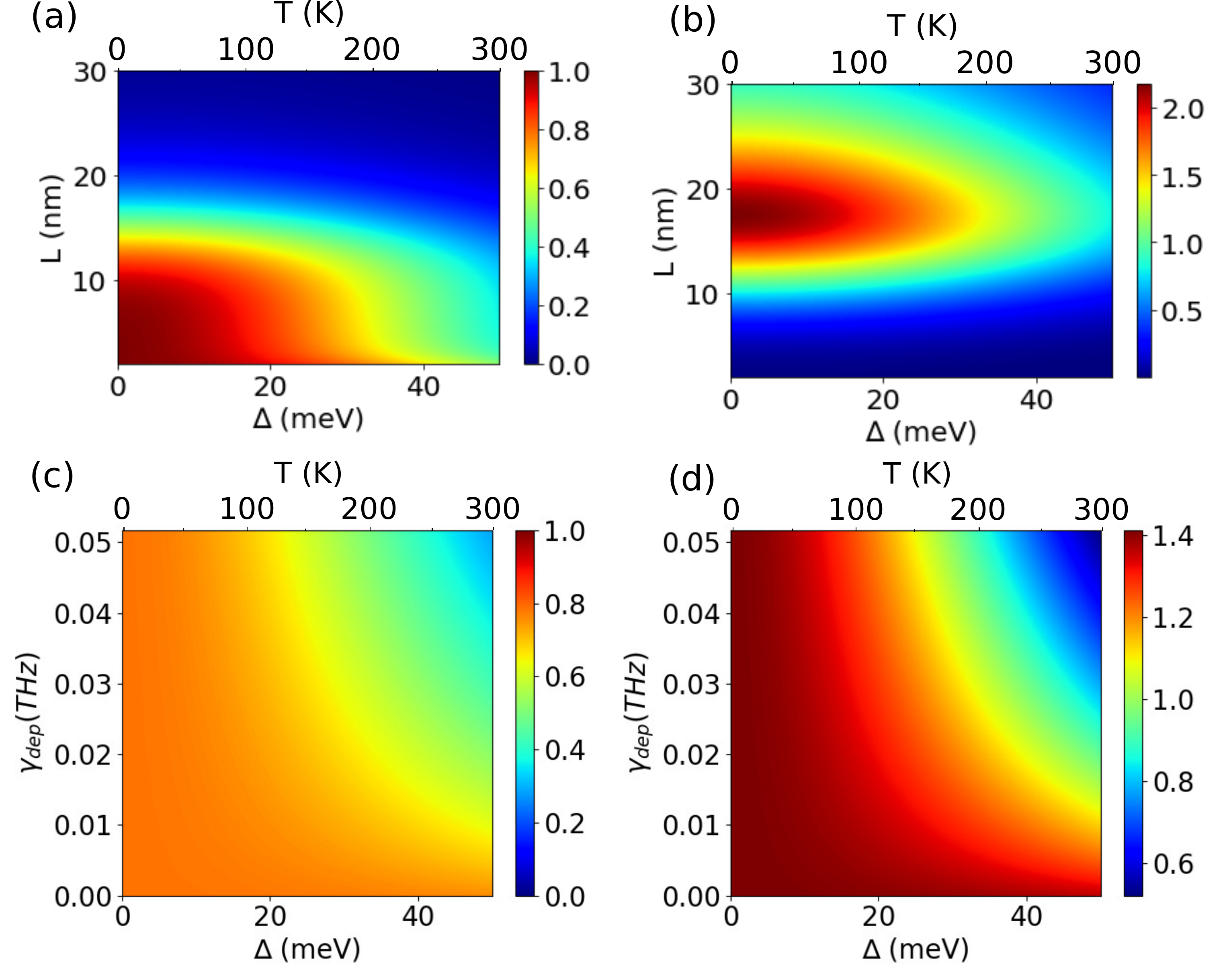}
 		\caption[]{(a) Contrast $C$ of the designed SET as a function of the anti-crossing energy $\Delta$ and the distance $L$ between the qubit and SET. (b) Same as (a), but for the differential conductance $g_m$  in unit of ($e^2/\hbar$).
 		(c) Contrast $C$ and (d) differential conductance $g_m$ as a function of $\Delta$ and of the dephasing rate $\gamma_{dep}$ with $L$=12.5~nm. 
 		$\Delta$ relates to the thermodynamic temperature $T=\Delta/2k_B$ limiting qubit operation.}
	\label{fig:6}
\end{figure}

\par The qubit is initialized by the application of an external electric field into one  of the two states $\ket{0}$ or $\ket{1}$. The system then evolves until it reaches a steady state at time $\sim1/\Gamma_R$ where $\Gamma_R$ is the escaping rate of the electrons to the drain (see Fig.~\ref{fig:5}(b)). The output current $I=2\pi e \Gamma_R \rho_R / \hbar$ is proportional to the population in the SET $\ket{R}$ state $\rho_R$. In what follows, we set  $\Gamma_R$ = $\Gamma_L$ = 0.12~THz in agreement with experimental studies~\cite{DQD}. To assess the measurement capabilities of the proposed SET we record its contrast $C$ and conductance $g_m$ during the qubit dynamics. 
Here, the contrast determines how well both qubit states can be distinguished. It is defined as
\begin{equation} \label{eq:contrast}
    C = \abs{I(1)-I(0)}/\abs{I(1)+I(0)},
\end{equation}
where $I(0)$ and $I(1)$ denote the output current when measuring a qubit in the $\ket{0}$ 
and $\ket{1}$ states, respectively.
The differential conductance quantifying the SET current changes caused by variations of the qubit states is defined as~\cite{SET}
 \begin{equation} \label{eq:conductance}
    g_m = C e [I(1)+I(0)] / \abs{ W_{T,R}-W_{B,R} }.
\end{equation}

The resulting contrast and differential conductance are plotted in Fig.~\ref{fig:6} as a function of the qubit anti-crossing energy $\Delta$ and distance $L$ between the qubit and SET with a characteristic time of $20/\Gamma_{R}$, including the aforementioned dissipative processes ($\gamma_{rel}=0.05$~THz based on low-temperature transport measurement~\cite{Stampfer2014}, while $\gamma_{dep}=2\gamma_{rel}$ to account for the fact that dephasing happens faster than relaxation). The bottom left corner of Fig.~\ref{fig:6}(a) shows the limiting range of $L$ (up to 13 nm) and $\Delta$ (up to 25 meV) to obtain high contrasts. Figure~\ref{fig:6}(b) also reports that $g_m$ decreases as a function of $\Delta$, but in a limited range of $L$ values between $12$ and $25$ nm.
 Combining both sub-plots, it clearly appears that the optimal conditions for qubit measurement are $L$=12.5~nm and $\Delta$ between 0 and 25~meV. The maximum $g_m$ value is higher than the experimental value reported in Ref.~\cite{Barthel2010} using a different SET setup. A widely accepted idea is that a large $\Delta$ is desired between the qubit states, in order to make them energetically distinguishable from each other. A necessary, but insufficient condition is that $\Delta$ must be larger than two times the thermal energy ($2k_B T$)  to suppress thermalization effects~\cite{Makhlin2001}. Therefore, a smaller $\Delta$ value necessarily entails a lower operation temperature of the qubit, as indicated by the upper abscissa axis in Fig~\ref{fig:6}. Here, we show that a normally functional qubit also imposes an upper-limit on $\Delta$, thus reducing the qubit design space.
 
\begin{figure}[htp]
	\centering
        \includegraphics[width=8cm]{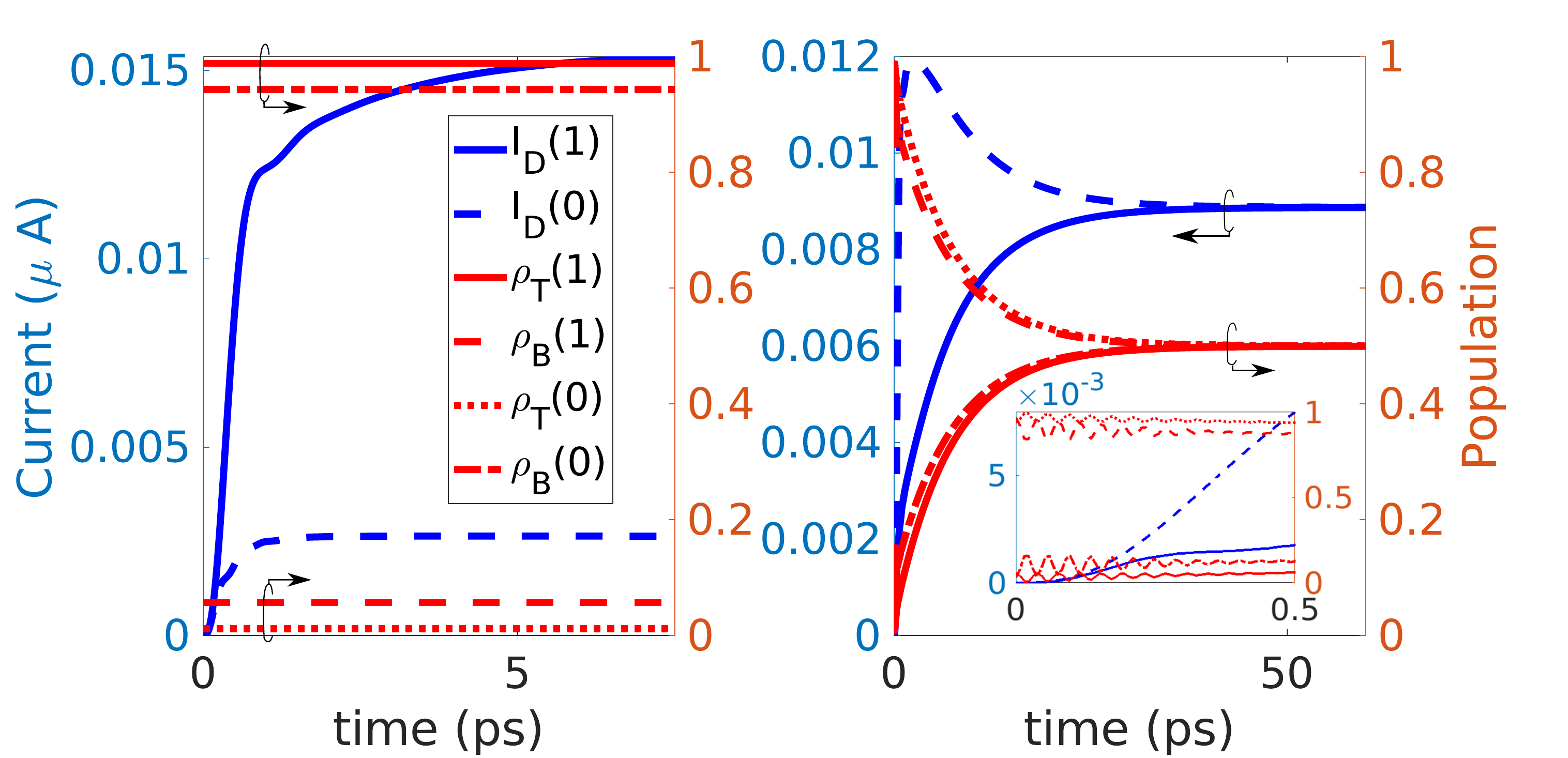}
 		\caption[]{(a) Time evolution of the drain current $I_D(0)$ and $I_D(1)$ flowing through a SET electrostatically coupled to a vdW qubit system with two logical states ($\ket{0}$ and $\ket{1}$) and of the top layer ($\rho_T(0)$ and $\rho_T(1)$) and bottom layer ($\rho_B(0)$ and $\rho_B(1)$) population for the BLG system. (b) Same as (a), but for the ZrS$_2$-HfS$_2$ system. The inset shows the short-time evolution after the beginning of measurement.}
	\label{fig:7}
\end{figure}

To examine the intrinsic feasibility of vdW charge qubits, we study $C$ and $g_m$ as a function of the dephasing rate and anti-crossing energy in Fig.~\ref{fig:6}(c) and (d). The distance between the SET and qubit is fixed to $12.5$ nm to ensure that both measurement quantities have high resolutions. It can be seen that the dephasing rate has a small impact on the degradation of $C$ and $g_m$ as long as $\Delta$ remains below 20 meV. However, when $\Delta$ exceeds 20 meV, $g_m$ and $C$ start to significantly decrease. This parameter region should therefore be avoided for the qubit design. 
This means that the anti-crossing energy is an intrinsically limiting factor, which suggests that the charge qubit should be defined with $\Delta \leq 20$ meV and the operating temperature lower than 120~K.

\begin{table}[b]
\begin{center}
\caption{\label{tab:table1}
Summary of the material properties and qubit measurement results ($C$ and $g_m$) for the BLG and ZrS$_2$-HfS$_2$ systems with $\vec{E}=-3~V/nm$, $\gamma_{dep}=2\gamma_{rel}=0.1$~THz. The $\Delta$ is the anti-crossing energy, whereas
$\vec{E}_{ac}$ denotes the electric field at the anti-crossing point when $t_{0,1}=b_{0,1}$.}
\begin{tabular}{@{}lcc@{}}
\toprule
    &BLG&ZrS$_2$ (T) HfS$_2$ (B)\\
\hline
 Contrast [\%] & 71 & 26\\
 $g_m$ [$\mu$S] & 102 & 35.7 \\
 $\Delta$ [meV] & 0.02 & 174  \\
 $\vec{E}_{ac}$ [V/nm] &  0  &  1.6   \\
\botrule
\end{tabular}
\end{center}
\end{table}

A typical time evolution of the drain current in this SET and the corresponding charge population in the qubit $\ket{T}$ ($\rho_T$) and $\ket{B}$ ($\rho_B$) states are presented in Fig.~\ref{fig:7} for BLG and
for ZrS$_2$-HfS$_2$. The BLG exhibits a much higher contrast of saturation current and almost no mixing. For ZrS$_2$-HfS$_2$, the strong $\Delta$ results in a fast qubit state mixing and consequently requires very rapid measurements within the picosecond time scale.   By reducing the strength of $\Delta$, for example by inserting an $h$BN monolayer between the vdW bilayers, the qubit measurement quality can be recovered. These results underline the fact that  high contrast  qubit measurements are intrinsically not possible if the charge qubit is characterized by a large $\Delta$.
Table I summaries the calculated qubit measurement results.

\section{Conclusion and Discussion}
\label{Discussion}
In this work, we proposed a vdW charge qubit architecture based on 2D vdW materials. Our results indicate that the qubit states can be readily controlled by external electric fields. The two-level anti-crossing energy, which is related to the vdW interlayer coupling strength, can significantly vary between different material systems. Our quantum transport simulations suggest that quantum dots can be confined by electrically gating the vdW bilayer and that the coupling between neighboring quantum dots can be tuned by the gate potential. Furthermore, we also proposed a single-electron transistor (SET) design based on a trilayer vdW heterostructure and capacitively coupled to the quantum dot to measure the charge qubit state. Our time evolution simulations of the coupled qubit and SET systems reveal that the measurement sensitivity and contrast are affected by the intrinsic anti-crossing energy and by the dephasing processes in the vdW heterostructure. Our study identifies the optimal design parameter space to simultaneously achieve a low decoherence of the qubit states and enhanced measurement contrast and conductance. 

In conclusion, our theory explains that upon the SET measurement the vdW-heterostructure qubit states rapidly degrade due to the large coupling intrinsically induced by anti-crossing energies higher than 20~meV.
This phenomenon was not captured by previous theoretical studies. Amoung the systems we studied, bilayer graphene is the one  maintaining the highest contrast between qubit states.
Our findings show that 2D vdW structures can provide highly integrable all-electronic universal qubits.  
Our design of the coupled vdW qubit and SET system allows for an enhanced sensitivity of the charge qubit, thus paving the way for a scalable and CMOS-compatible quantum computing platform.

\appendix 

\section{Ab initio quantum transport simulation}
\label{Methodology1}
\par Following Ref.~\cite{ChargeQubit}, a charge qubit can be defined in 2D bilayer materials through the superposition of a top
($\ket{T}$) and bottom ($\ket{B}$) orbital state, whose properties depend on a vertically applied electric field ($\Vec{E}$). Since each
local orbital occupies a different electrically-induced energy state, two distinct energy levels with different weights ($t(\Vec{E})$ and $b(\Vec{E})$) can be obtained (Fig.~1(a)). They are used as building blocks for the Bloch sphere and 
its  $\ket{0}$ and $\ket{1}$ states, which form the two-level quantum system shown in Fig.~1(b). The latter is simulated at the \textit{ab initio} level.

\par Through the QUANTUM ESPRESSO DFT package \cite{QE}, the electronic structure of 2D bilayer materials is first obtained. The generalized gradient approximation of Perdew, Burke, and Ernzerhof (PBE)~\cite{PBE} is used as exchange and correlation functional, while the vdW interactions are taken into account with the DFT-D2 scheme of Grimme~\cite{GrimmeD2}. The plane-wave DFT Hamiltonian is then converted into a maximally-localised Wannier function (MLWF) basis with the Wannier90 code~\cite{wannier90}. These results are scaled up \cite{Szabo2019electron} to construct the real-space device Hamiltonian and to perform realistic qubit device simulations based on the Non-equilibrium Green's Function (NEGF) formalism. The following equation is solved for the retarded Green's function $\mathbf{G^R}$,
\begin{equation}
    \left[ E \mathbf{I} - \mathbf{H}_{\rm MLWF} - \mathbf{\Sigma}_{\rm S/D}^{\rm R} \right] \mathbf{G}^{\rm R} = \mathbf{I} ,
\end{equation}
where $E$ is the electron energy, $\mathbf{I}$ the identity matrix, 
$\mathbf{H}_{\rm MLWF}$ the Wannier Hamiltonian matrix of the 2D bilayer graphene in the MLWF basis, and $\mathbf{\Sigma}_{\rm S/D}^{\rm R}$ the retarded boundary self-energy accounting for the source and drain semi-infinite contacts. The observables of the system can be obtained from the retarded Green's function  $\mathbf{G}^{\rm R}$. The periodic boundary condition is imposed along the $y$ direction as shown in Fig.~\ref{fig:3}(a).
By including an electrostatically-induced confinement
potential into $\mathbf{H}_{\rm MLWF}$, a wide range of QD systems can be realized and their operation verified by inspecting the local density-of-states (LDOS). The electrical current  of the device is calculated with the Landauer-B\"uttiker formula from the transmission probability $T(E)$ between two contacts~\cite{Landauer,Buttiker}
\begin{equation}
    I = - \frac{e}{\hbar}  \int  \frac{dE}{2\pi} T(E) [f(E,E_{fL}) - f(E,E_{fR})],
\end{equation}
with the elementary charge $e$. The two contacts are characterized by their Fermi distribution function $f$ and Fermi levels $E_{fL}$ and $E_{fR}$, whose difference depends on the applied source-to-drain voltage $E_{fL}-E_{fR}=eV_{DS}$. The conductance $G$ is calculated as $dI/dV_{DS}$.\\

\section{Evaluation of time-dependent Lindblad equation for qubit readout}
\label{Methodology2}
\par The dynamics of the charge qubits is simulated with the QuTiP \cite{Qutip} package, which 
solves the Lindblad equation~\cite{SET},
\begin{equation} \label{eq:time_evolve}
\begin{split}
    \frac{d \hat{\rho}}{dt} = &-i\hbar \left[ \hat{H}_{\rm Q} + \hat{H}_{\rm SET} + \hat{H}_{\rm Q-SET}, \hat{\rho}   \right]  \\
    & + \mathcal{C} + \Gamma_R \ket{vac_R}\bra{R} + \Gamma_L\ket{L}\bra{vac_L},\\     
\end{split}
\end{equation}
where the qubit Hamiltonian ($H_{\rm Q}$) is a two-level quantum system constructed from the electronic structure of a 2D bilayer material, while a three-level QD is used to represent the SET Hamiltonian ($H_{\rm SET}$). The qubit and SET parts are only electrostatically coupled through the electron-electron Coulomb interaction $H_{\rm Q-SET}$, assuming no tunneling between them. The dissipative processes including relaxation and dephasing are described by the collapse operators $\mathcal{C}$ in Eq.~(\ref{eq:c_ops}). The meaning and definition of each operator are given in Section~\ref{QubitReadout}. 
An arbitrary state as in Eq.~(\ref{eq:state}) is evolved by integrating the set of ordinary differential equations that define Eq.~(\ref{eq:time_evolve}). The expectation value of the populations in the SET are recorded at each time step by means of the operators $\rho_L=\ket{L}\bra{L}$, $\rho_R=\ket{R}\bra{R}$, and $\rho_C=\ket{C}\bra{C}$. The output at the last time step is used in Eqs.~(\ref{eq:contrast}) and (\ref{eq:conductance}) to asses the contrast and conductance of the proposed SET.

\begin{acknowledgments}

This work was supported by the NCCR MARVEL and the NCCR SPIN of the Swiss National Science Foundation (SNSF), by SNSF under Grant No.~175479 (ABIME), by the Marie Skłodowska-Curie Grant No. 885893, and by a grant from the Swiss National Supercomputing Centre (CSCS) under Project s1119.

\end{acknowledgments}

\bibliography{main}

\end{document}